\documentclass[sigconf,nonacm]{acmart}

\usepackage{listings}
\usepackage{xcolor}
\usepackage[T1]{fontenc}

\definecolor{swiftkeyword}{RGB}{170,13,145}     
\definecolor{swiftcomment}{RGB}{0,116,0}        
\definecolor{swiftstring}{RGB}{196,26,22}       
\definecolor{swiftnumber}{RGB}{28,0,207}        
\definecolor{swiftgray}{RGB}{128,128,128}       

\lstdefinelanguage{Swift}{
  morekeywords={func, struct, enum, case, protocol, associatedtype, let, var, if, else, for, in, repeat, each, guard, return, throw, throws, do, catch, print, any, true, false, try},
  sensitive=true,
  morecomment=[l]{//},
  morestring=[b]",
  morestring=[b]"""
}

\usepackage{graphicx}
\usepackage{tikz}
\usetikzlibrary{arrows.meta, positioning}

\AtBeginDocument{%
  }

\setcopyright{none}
\copyrightyear{2026}
\acmYear{2026}
\acmDOI{XXXXXXX.XXXXXXX}
\acmConference[Conference acronym 'XX]{Make sure to enter the correct
  conference title from your rights confirmation email}{June 03--05,
  2018}{Woodstock, NY}
\acmISBN{978-1-4503-XXXX-X/2018/06}

\begin{document}

\title{Pack Iteration in Swift:\\Ordinary Control Flow for Variadic Generics}

\author{Serafima Nerush}
\orcid{0009-0000-9334-1880}
\authornote{Both authors were with Reed College, Portland, OR, USA, at the time this work was performed.}
\affiliation{%
  \institution{}
  \city{}
  \state{}
  \country{}
}

\author{Eitan Frachtenberg}
\orcid{0000-0002-3709-1829}
\affiliation{%
  \institution{}
  \city{}
  \state{}
  \country{}
}

\renewcommand{\shortauthors}{Nerush and Frachtenberg}

\begin{abstract}
Variadic generics are a powerful tool for type-safe meta-programming. Yet in most widely used languages, they remain an 
``expert-only'' feature due to their reliance on complex 
patterns such as recursive decomposition
or expansion expressions that do not compose naturally with 
ordinary control flow. In C++, for example, accessing elements 
of parameter packs has traditionally relied on unintuitive 
recursive patterns that ``peel off'' elements.

This paper presents \emph{Pack Iteration}, a feature introduced in Swift 6.0 that allows developers to iterate over parameter packs using a familiar, imperative \texttt{for-in} loop.
By treating pack expansion as a first-class source for iteration, Swift bridges the gap between high-level expressiveness and advanced generic programming.

We detail the design and implementation of this feature within the Swift compiler, focusing on the challenges of bridging static type-checking in the constraint system with dynamic execution in the Swift Intermediate Language.

Unlike traditional models that expand packs at compile time, Swift's implementation supports on-demand evaluation, enabling efficient dynamic iteration and short-circuiting control flow.
Our empirical evaluation confirms that pack iteration
provides performance comparable to---and sometimes significantly better than---the complex workarounds previously required.

More broadly, pack iteration shows how
advanced generic programming can be made to feel like ordinary
programming while preserving type safety and efficient 
execution.
\end{abstract}



\maketitle

\section{Introduction}

Variadic generics allow functions and data types to abstract over an
arbitrary number of type arguments while preserving the identity of
each individual type. This makes them well suited to programming patterns
in which arity is not known in advance, such as tuple operations and other
interfaces over heterogeneous collections of values. By lifting
fixed-arity generic abstractions to variable arity, variadic generics
enable these patterns to be expressed in a type-safe manner.

Despite this expressive power, variadic generics are viewed as a
feature geared towards more advanced programmers in the mainstream programming
languages because their typical syntax and implementation are complex.
C++, for example, implements the concept of variadic generics 
in the form of variadic templates~\cite{Gregor2007Variadic}.

Traditionally, operating on parameter packs in C++ has relied on template meta-programming techniques that utilize recursive instantiation to traverse types at compile-time~\cite{veldhuizen2006tradeoffs}.
Although highly performant, this model often leads to code that is difficult to reason about and maintain.
In order for the programmer to operate on the parameter pack, they have to
recursively ``peel off'' elements and continue to recurse into the remainder of
the pack.

The parameter packs feature in Swift has also exposed a
similar usability problem: parameter packs could only be
expanded into an expression. Consequently, using parameter
packs was hard and unintuitive without a natural way
for a programmer to operate on their elements.

The difficulty is not merely syntactic. Inline 
expansion expressions lacked control flow
support, forcing programmers to use unnatural workarounds like stopping iteration
by throwing an error in an additional local throwing function.
Swift positions itself as a language that is both approachable for beginners and powerful enough for systems-level tasks.
This design philosophy---balancing type safety with high-level expressiveness---is central to the language's evolution~\cite{Apple2017, saeta2021swift}.

This paper posits that the problem is not inherent to variadic generics
themselves, but to the programming model through which they are exposed.
We present \emph{Pack Iteration}, a novel model in which a pack expansion
expression can serve as a source of ordinary iteration.
Concretely, we describe its realization in Swift through SE-0408\footnote{\url{https://github.com/swiftlang/swift-evolution/blob/main/proposals/0408-pack-iteration.md}}, which
extends the familiar \texttt{for-in} loop to iterate over value packs.
This design allows programmers to bind pack elements directly to a local
variable declared inside of a \texttt{for-in} loop. Moreover, it allows
the use of ordinary control-flow constructs such as \texttt{guard},
\texttt{break}, and \texttt{continue}. Together, this enables developers
to express complex pack-processing operations in an idiomatic way,
making variadic generics in Swift more approachable.
By lowering the barrier to variadic generics, Pack Iteration makes this kind of meta-programming practical for everyday Swift code. Instead of thinking in terms of recursive decomposition, programmers can process a pack with the same sequential control flow they use elsewhere.

Pack iteration is also interesting from an implementation perspective.
Unlike traditional approaches that expand packs entirely at compile time
(such as C++), Swift supports dynamic, on-demand evaluation of pack
elements during iteration. This enables short-circuiting behavior
without requiring all elements of the pack to be eagerly processed.
We argue that Swift is the first widely used programming language
to support explicit pack iteration with an ordinary \texttt{for-in}
loop.

The contributions of this paper are therefore twofold. First, it presents
pack iteration as a language design paradigm that makes variadic
generics substantially more approachable and usable in practice.
Second, it describes and evaluates the first full implementation of this 
design in the Swift programming language, where pack iteration is
implemented through dynamic, on-demand execution rather than pure compile-time expansion.

\section{Background and Related Work}

To motivate parameter pack iteration, we first review the aspects of
Swift's generics system that are most relevant to this feature. We also
compare Swift with other widely used languages to
highlight the distinctiveness of Swift's parameter packs and the design
space for operating on them.

\subsection{Generics}

Mainstream programming languages like C++, Swift, and Java employ distinct models of generic programming rooted in their respective implementation strategies~\cite{ghosh2004generics}.

C++ supports generic programming in the form of function or class templates~\cite{gregor2006concepts}. The implementation of templates relies on \emph{algorithm specialization}, meaning generics are purely a compile-time feature. They
exist in AST but cannot be lowered beyond that, thus having no runtime presence. Although widely used, templates in C++ cannot be compiled,
or even type-checked, in isolation of their uses~\cite{Jarvi2006}. This
implementation model results in a poor type checking and diagnostic experience for the programmer~\cite{chen2020c++}.

In Java, the implementation of generics is based on the \emph{type erasure}
model, where the compiler erases the generic type and replaces it with
the bounding type \texttt{Object}, inserting casts into the actual places of usage~\cite{radenski2008java}.
This yields a \emph{homogeneous} translation, as generic definitions are compiled
once rather than specialized for each use~\cite{1998bracha}. Compared to
C++, where this process is
\emph{heterogeneous} and type-safe~\cite{ghosh2004generics}, Java discards most
generic type information at runtime.

Swift differs from both C++ and Java in that it
combines separate type-checking with a runtime model
for unspecialized generic code. As described by \citet{Pestov2022}, Swift's generic implementation is
guided by four principles. First, generic declarations must be type-checked
\emph{independently} of their uses, so specialization is not
required. Second, generic libraries must support resilient evolution
without forcing client recompilation. Third, the actual layouts of
generic types are determined by their concrete substitutions, and
fields of generic parameters are stored inline. Finally, the use of
generics should incur runtime cost only when compile-time type
information is unavailable. 

To satisfy these constraints, Swift uses the generic signature as the interface
between a generic declaration and its callers, and for each generic
argument from the caller, generates \emph{runtime type metadata}, which
helps manipulate values without knowing their layout. The \emph{runtime type metadata}
is then used for the \emph{runtime type layout} of the generic type, without requiring boxing or indirection. 
Specialization that eliminates this overhead may be introduced as an
optimizer step, but unlike in C++, it is
not required for correctness or separate compilation;
and unlike in Java, this model does not depend on type erasure at runtime.

\subsection{Variadic Generics}

Variadic generics is a concept that builds on top of the generics implementation
model in programming languages, in a way that allows to express a declaration
to be generic over any number of types, while still being able to know what
exactly these types are. This makes them especially useful for APIs
where the arity is not known in advance, such as tuple operations and
interfaces over heterogeneous collections of values. In practice,
variadic generics are realized through \emph{parameter packs}, which
represent a ``list'' of types or values.

\subsubsection{Variadic Templates in C++} \label{sec:variadicC++}

The implementation of variadic generics in C++, variadic templates, was
first introduced by \citet{Gregor2007Variadic}. Prior to their adoption, programmers resorted to workarounds
with regular templates that \emph{emulated} the variadic version of them via
manipulating the preprocessor. These techniques were not efficient and
introduced several limitations that made them difficult to use.

Crucially, Gregor et al.'s work introduced the concept
of a \emph{parameter pack}, a special type parameter that can hold any number
of types. It is \emph{unpacked} with the special ellipsis
operator at usage sites, ``expanding'' or substituting
into the corresponding type parameters.

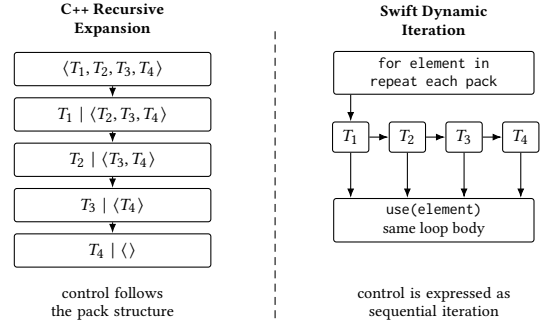
\begin{figure}[t]
\centering
\begin{tikzpicture}[
  font=\scriptsize,
  stage/.style={
    draw,
    rounded corners=1pt,
    minimum width=2.60cm,
    minimum height=0.44cm,
    align=center
  },
  item/.style={
    draw,
    rounded corners=1pt,
    minimum width=0.48cm,
    minimum height=0.40cm,
    align=center
  },
  loopbody/.style={
    draw,
    rounded corners=1pt,
    minimum width=2.65cm,
    minimum height=0.50cm,
    align=center
  },
  note/.style={
    align=center,
    text width=2.75cm
  },
  arrow/.style={-{Latex[length=1.25mm]}, thin},
  every node/.style={inner sep=2pt}
]

\node[font=\bfseries\scriptsize, align=center] at (0,0)
  {C++ Recursive\\Expansion};
\node[font=\bfseries\scriptsize, align=center] at (4.25,0)
  {Swift Dynamic\\Iteration};

\node[stage] (c0) at (0,-0.65) {$\langle T_1,T_2,T_3,T_4\rangle$};
\node[stage] (c1) at (0,-1.25) {$T_1 \mid \langle T_2,T_3,T_4\rangle$};
\node[stage] (c2) at (0,-1.85) {$T_2 \mid \langle T_3,T_4\rangle$};
\node[stage] (c3) at (0,-2.45) {$T_3 \mid \langle T_4\rangle$};
\node[stage] (c4) at (0,-3.05) {$T_4 \mid \langle\rangle$};

\draw[arrow] (c0) -- (c1);
\draw[arrow] (c1) -- (c2);
\draw[arrow] (c2) -- (c3);
\draw[arrow] (c3) -- (c4);

\node[note] at (0,-3.75)
  {control follows the pack structure};

\node[loopbody] (s0) at (4.25,-0.70)
  {\texttt{for element in}\\\texttt{repeat each pack}};

\node[item] (s1) at (3.15,-1.55) {$T_1$};
\node[item] (s2) at (3.90,-1.55) {$T_2$};
\node[item] (s3) at (4.65,-1.55) {$T_3$};
\node[item] (s4) at (5.40,-1.55) {$T_4$};

\draw[arrow] (s0.south -| s1.north) -- (s1.north);
\draw[arrow] (s1.east) -- (s2.west);
\draw[arrow] (s2.east) -- (s3.west);
\draw[arrow] (s3.east) -- (s4.west);

\node[loopbody] (body) at (4.25,-2.65)
  {\texttt{use(element)}\\same loop body};
\foreach \s in {s1,s2,s3,s4}
  \draw[arrow] (\s.south) -- (\s.south |- body.north);

\node[note] at (4.25,-3.75)
  {control is expressed as sequential iteration};

\draw[densely dashed] (2.15,-0.15) -- (2.15,-3.95);

\end{tikzpicture}

\caption{Two models for processing a parameter pack. C++ variadic-template
idioms decompose the pack into a head and tail, yielding a recursive chain of
template specializations. Swift pack iteration treats the pack as a sequence of
opened elements processed by a single loop body.}
\label{fig:pack-processing-models}
\end{figure}

The implementation model of variadic templates described by~\citet{Gregor2007Variadic} is an additive change on top of the
existing C++ templates, where at compile-time, specializations
are produced. Therefore, parameter-pack manipulation is only possible
in terms of expansion rather than traversal. Traditional idioms expose this model
through recursive template decomposition, where one element of a pack is
``peeled off'' and the remainder is processed recursively. Although this style can
be efficient in practice, it also makes many pack-processing tasks difficult to
read and express directly. Figure~\ref{fig:pack-processing-models} contrasts this recursive
decomposition model with Swift's sequential pack-iteration model.

We now turn to Swift's parameter packs and the limitations that
motivated adding pack iteration as a
first-class operation.

\subsection{Parameter Packs in Swift}

Similarly to C++, Swift supports
\emph{type} and \emph{value parameter packs}. The immediate motivation
was to eliminate families of fixed-arity overloads while preserving type
information, allowing APIs such as tuple operations to be expressed by
a single generic declaration rather than by defining $n$ overloads, each
accepting $0..(n-1)$ generic arguments to support tuples of $n-1$ elements \cite{SE0393}. For example, tuple comparison can be written once for tuples
of \emph{arbitrary} length:

\begin{lstlisting}
func == <each Element: Equatable>(
  lhs: (repeat each Element),
  rhs: (repeat each Element)
) -> Bool
\end{lstlisting}

The special \texttt{each} syntax above applied to a
generic argument of a function turns a type parameter
into a \emph{type parameter pack}, meaning that it can accept, or \emph{expand} into,
any number of generic arguments. Moreover, packs support
protocol constraints—in the above example, the list of types
\texttt{Element} expands to needs to conform to the \texttt{Equatable} protocol. This is type-checked against the call site, and within the body of the function the programmer
can use the values of type \texttt{each Element} in contexts
requiring comparison ability, just like the tuple operator above.

Analogously to the ellipsis in C++ pack expansion, the expression to
which Swift's \texttt{repeat} keyword is applied is called the
\emph{repetition pattern}. At runtime,
this pattern is repeated for each element in the substituted pack and the resulting
types or values are expanded into the list provided by the surrounding context \cite{SE0393}.
In this case, if \texttt{each Element} is 
substituted into \texttt{\{Int, Bool, String\}}, then this pattern will be 
repeated for unlabeled elements of 
\texttt{lhs} and \texttt{rhs} tuples, 
producing two tuples with elements of type 
\texttt{(Int, Bool, String)}. 
Additionally, if \texttt{repeat each lhs} 
is used in the body of the function, it 
will be called a \textit{value pack 
expansion}: \texttt{lhs} is a tuple whose 
elements form a value pack, so the pattern 
is repeated for each of the element 
\textit{values} of \texttt{lhs}, 
corresponding to each of the types in the 
\textit{type} parameter pack.

The motivation for this paper stems from the limitation where values of a given
parameter pack in Swift can only be
accessed from a \textit{pack expansion expression}---the expression that follows the \texttt{repeat} keyword.
This design requires the programmer
to express traversal over a value 
parameter pack by composing \texttt{repeat each} expressions with a limited set of 
accepted positions, such as function 
argument lists and tuple literals.

\section{Pack Iteration in Swift}

In this section, we detail the primary motivation and implementation
of \emph{parameter-pack iteration} in the Swift programming language.

\subsection{The Problem}

As discussed earlier, the support of variadic generics
in Swift was limited in a way
that any processing operation on \emph{value} parameter packs
required structuring the code around expressions, forcing the
programmer into unnatural patterns.

Expanding a repetition pattern within a pack expansion expression forces eager evaluation for every element, making short-circuiting behavior difficult to express.
One approach requires putting code involving statements into a \emph{throwing} function to enable such short-circuiting behavior. Then, the programmer
can catch the error in a \texttt{do}/\texttt{catch} block to return, making this pattern very unintuitive.
    
Consider the following implementation of the \texttt{==} operator over tuples of arbitrary length that demonstrates the above pattern:

\begin{lstlisting}
struct NotEqual: Error {}

func == <each Element: Equatable>(
    lhs: (repeat each Element),
    rhs: (repeat each Element)
) -> Bool {
  // Local throwing function for operating
  // over each element of a pack expansion.
  func isEqual<T: Equatable>(
    _ left: T,
    _ right: T
  ) throws {
    if left == right {
      return
    }
    
    throw NotEqual()
  }

  // Do-catch statement for short-circuiting as
  // soon as two tuple elements are not equal.
  do {
    repeat try isEqual(each lhs, each rhs)
  } catch {
    return false
  }

  return true
}
\end{lstlisting}

In the above code, the programmer can only stop ``iteration'' and
return \texttt{false} when the \texttt{NotEqual} error is thrown.
The \texttt{isEqual} function performs the comparison and throws if the sides are not equal.

The lack of first-class short-circuiting support and expression-only semantics made variadic generics
in Swift an advanced feature, accessible only to expert programmers.

\subsection{Introducing Pack Iteration}

In this paper, we posit that an ordinary \texttt{for-in}
loop in Swift can provide first-class support for
iterating over a value parameter pack. Under this model,
which was proposed in \emph{SE-0408: Pack Iteration} \cite{nerush0408} Swift
language proposal, the implementation of the tuple equality operator from the previous section turns into a simple \texttt{for-in} loop:

\begin{lstlisting}
func == <each Element: Equatable>(
  lhs: (repeat each Element),
  rhs: (repeat each Element)
) -> Bool {
  for (left, right) in repeat (each lhs, each rhs) {
    guard left == right else { return false }
  }
  return true
}
\end{lstlisting}

In Swift, the \texttt{guard} statement provides early exit control flow: if the condition evaluates to false, the mandatory \texttt{else} block executes (here returning false), avoiding deeply nested if blocks.
Note that in the above code, the \texttt{NotEqual} struct
is no longer needed, as well as the nested \texttt{isEqual}
function along with the \texttt{do}/\texttt{catch} block. 

Instead, the programmer can simply utilize
the \texttt{for-in} loop to iterate over the
elements from both value pack tuples pairwise
by assigning the value of the pack element to the
local variable on each iteration. The programmer
can escape the function by returning \texttt{false}
 from the function instead of throwing an error.

Crucially, the \texttt{for-in} loop unlocks operations over value packs that were previously impossible; it allows for the creation of a local tuple \texttt{(left, right)}, bound to a pair of
elements from \texttt{lhs} and \texttt{rhs} value pack tuples,
respectively. This is powerful as it allows the programmer
to express advanced control flow in the body of the loop.
The Swift compiler proves that because \texttt{lhs} and
\texttt{rhs} are of the same type, so are \texttt{left} and \texttt{right}. The \texttt{Equatable} requirement declared
on the type pack allows comparing the local variables
with the \texttt{==} operator.

\subsubsection{Language Design}
\label{subsec:design}

Now, in addition to expressions that conform to the
\texttt{Sequence} protocol, the source of
a \texttt{for-in} loop may also be a
pack expansion expression.

Consider the following reduced example of pack iteration:

\begin{lstlisting}
func iterate<each Element>(
  over element: repeat each Element
) {
  for el in repeat each element { }
}
\end{lstlisting}

On the $i$\textsuperscript{th} iteration, the type of
\texttt{el} is the $i$\textsuperscript{th} type
parameter in the \texttt{Element} type parameter pack.
The value of \texttt{el} is the $i$\textsuperscript{th}
value parameter in the \texttt{element} value parameter pack.

What is the type of the local variable \texttt{el}?
Generally, its type is the pattern type of \texttt{repeat each element} with each captured type parameter pack replaced with
an implicit scalar type parameter with matching requirements.
In this case, the pattern type of \texttt{repeat each element}
is \texttt{each Element}, and, therefore, the type of \texttt{el}
is a scalar type parameter with no requirements.
For clarity, we refer to this scalar type parameter as \texttt{Element'}. 

As such, if the \texttt{iterate} function is called with
\texttt{each Element} bound to the type pack
\texttt{\{Int, String, Bool\}}, within the body of the
\texttt{for-in} loop, the type \texttt{Element'} will be
substituted for \texttt{Int}, then \texttt{String}, 
and finally \texttt{Bool}.

If the type parameter packs captured by the pack
expansion pattern contain requirements, the scalar
type parameter in the loop body will have the same
requirements:

\begin{lstlisting}
struct Generic<T> {}

protocol P {}

func iterate<each Element: P>(
  _ element: repeat each Element
) {
  for x in repeat Generic<each Element>() {
    // the type of 'x' is <Element': P> Generic<Element'>
  }
}
\end{lstlisting}

In the above code, the pattern type of the pack
expansion is \texttt{Generic<each Element>}, where
\texttt{each Element: P}, thus the type of local
variable \texttt{x} is \texttt{Generic<Element'>}, where
\texttt{Element'} is a scalar type parameter
with the requirement \texttt{Element': P}.

As mentioned previously, pack iteration gets the already existing \texttt{for-in}
loop functionality, naturally composing with the
patterns most programmers are already used to.

First, like regular \texttt{for-in} loops,
\texttt{for-in} loops over pack expansions can
pattern match over the element of the value pack:

\begin{lstlisting}
enum E<T> {
  case one(T)
  case two
}

func iterate<each Element>(
  over element: repeat E<each Element>
) {
  for case .one(let value) in repeat each element {
    // 'value' has type <Element'> Element'
  }
}
\end{lstlisting}

Here, for case applies algebraic data type (ADT) pattern matching to the enum payload: each iteration inspects the current pack element and executes the loop body only if the element matches the .one case, binding its payload to value.
In the above code, the repetition pattern
\texttt{each element} in the source of the
for-in loop is evaluated once at each iteration
(instead of $n$ times eagerly), where $n$ is the length
of the packs captured by the pattern. In other words,
if $p_i$ is the pattern expression at the $i$\textsuperscript{th} iteration, and control flow exits the loop at
iteration $i$, then $p_j$ is not evaluated for $i < j \le n$.
The following code example demonstrates this:

\begin{lstlisting}
func printAndReturn<Value>(
  _ value: Value
) -> Value {
  print("Evaluated pack element value \(value)")
  return value
}

func iterate<each T>(_ t: repeat each T) {
  var i = 0
  for value in repeat printAndReturn(each t) {
    print("Evaluating loop iteration \(i)")
    if i == 1 { 
      break 
    } else {
      i += 1
    }
  }
  
  print("Done iterating")
}

iterate(1, "hello", true)
\end{lstlisting}

The above code prints the following:

\begin{lstlisting}[language=]
Evaluated pack element value 1
Evaluating loop iteration 0
Evaluated pack element value hello
Evaluating loop iteration 1
Done iterating
\end{lstlisting}

The next example demonstrates a more advanced usage of
variadic generics, greatly simplified by pack iteration.

Consider the following protocol:
\begin{lstlisting}
protocol ValueProducer {
  associatedtype Value: Codable
  func evaluate() -> Value
}
\end{lstlisting}

The \texttt{ValueProducer} protocol above
requires an \texttt{evaluate} method whose return type,
\texttt{Value}, is constrained to conform to the \texttt{Codable} protocol.

Now consider a task where a parameter pack of values of type
\texttt{Result<V, E>}, where \texttt{V: ValueProducer} and \texttt{E: Error}, needs to be filtered into an array with only the values produced by its \texttt{.success} elements.

Pack iteration lets the programmer accomplish this with the
\texttt{for case} pattern:

\begin{lstlisting}
func evaluateAll<
    each V: ValueProducer, each E: Error
>(result: repeat Result<each V, each E>
) -> [any Codable] {
  var evaluated: [any Codable] = []
  for case .success(let producer) in repeat each result {
    evaluated.append(producer.evaluate())
  }

  return evaluated
}
\end{lstlisting}

This example shows why variadic generics are needed here. If \texttt{evaluateAll} were implemented over a heterogeneous collection such as \texttt{[Any]}, the compiler would lose the static conformance of each element to \texttt{ValueProducer}, so calling \texttt{producer.evaluate()} would require runtime downcasting. Pack iteration keeps each element's type and protocol conformances statically known across loop iterations, while the loop itself remains ordinary imperative code.

Let us also note the signature of the \texttt{evaluateAll} function.
In the generic parameter list, it declares two type parameter packs:
\texttt{each V: ValueProducer}, and \texttt{each E: Error},
constraining each element of them to \texttt{ValueProducer} and
\texttt{Error} protocols, respectively. The argument value
pack \texttt{result} of heterogeneous types
is then composed of these generic types,
expanding them at run time and guaranteeing the same lengths
at compile time.

The \texttt{for case} ensures that only
\texttt{Result.success} values are processed.
Within the body of the \texttt{for} loop, pack iteration
binds each value producer to a local variable,
\texttt{producer}, allowing the result of
\texttt{producer.evaluate()} to be appended to the
final array.

\subsection{Implementation Details}

This section documents the relevant implementation details of
\emph{Pack Iteration} in Swift.

\subsubsection{Overview}

The main implementation challenge goes beyond merely supporting new syntax; it requires ensuring Pack Iteration provides semantics similar to an ordinary
\texttt{Sequence}-conforming type while being a
completely distinct language construct.

To tackle this challenge, we have modified every step
of the Swift compiler\footnote{\url{https://www.swift.org/documentation/swift-compiler/}}, from semantic analysis to
generating the Swift intermediate language (SIL).

Semantic analysis (referred to as \textbf{Sema}) is a stage
of the Swift compiler responsible for transforming the parsed
abstract syntax tree (AST) into a
well-formed, fully-type-checked form of the AST,
emitting warnings or errors for semantic problems in
the source code.
In Sema, the compiler must recognize
that the source of a \texttt{for-in} statement
may be either a \texttt{Sequence} or a pack expansion expression,
and it must determine the type of the loop-bound
local variables accordingly. 

The compiler then \emph{lowers} the resulting AST to
SIL, a high-level, Swift-specific intermediate language
suitable for further analysis and optimization of the source
code. This SIL generation process is referred to as \textbf{SILGen}. In this stage, our goal was to ensure the
compiler preserves the control-flow behavior of a loop while
evaluating the pack elements incrementally rather than
expanding all of them eagerly.

\subsubsection{Semantic Analysis}

Before \emph{Pack Iteration}, semantic analysis performed
type-checking of the \texttt{for-in} loop under the assumption that it
denoted a value of a \texttt{Sequence}-conforming type. Under \emph{Pack Iteration},
the source of the loop may instead be a \texttt{PackExpansionExpr}, where
the element type is derived from the expansion pattern rather than by
the \texttt{Sequence} protocol. Consequently, the compiler must distinguish
between two semantic cases for the source of \texttt{for-in} statements: iterating over a sequence and a value pack.

We explicitly codified this distinction within the loop-specific bookkeeping by
the constraint system. We generalized the existing representation of \texttt{for-in}
statement information from a single sequence-specific record
into a tagged union with separate cases for sequence iteration and pack
iteration.

\begin{lstlisting}[language=C++]
struct ForEachStmtInfo : TaggedUnion<
  SequenceIterationInfo, PackIterationInfo >
{
    using TaggedUnion::TaggedUnion;
};
\end{lstlisting}

We found that \texttt{PackIterationInfo} only needs to store the type
of the pattern that matches the elements in the pack expansion for type-checking.

Pack iteration introduces a new typing context inside the loop body. Each
iteration conceptually opens one element of the surrounding pack, so
declarations inside of the loop need to have a special generic environment
which encodes the mapping from the pack types in the outer context to element
types inside the loop. Therefore, we have modified
Swift's variable declaration representation to store
this \emph{opened element environment}, allowing
declarations introduced in the scope of the loop to be
correctly type-checked using the corresponding generic
environment.

The core of the work then falls into the two main phases
of the constraint system, \texttt{CSGen} and \texttt{CSApply}. The 
constraint system is part of the semantic
analysis, and it first generates constraints
for the AST, solves them, and,
finally, applies the solution back to the syntax tree.

For pack iteration, both \texttt{CSGen} and \texttt{CSApply} had to be extended with pack-specific
overloads of the existing \texttt{for-in} machinery.

Specifically, in constraint generation, we first
obtain the element type from the pattern type of
the pack expansion. We then generate constraints for the expansion as a syntactic target, query the constraint system for the type of the expansion's pattern expression, and finally add a \emph{conversion} constraint between the element type and the pattern type. This constraint gives
Sema the necessary information to reason about
the typing rules for the local variable on each
iteration of the loop and therefore makes it possible
to prove that the user-written loop pattern must be
sound with the usage of the local variable. The use
of the \emph{conversion} constraint in conjunction
with the derived element type makes the type-checking
of the value pack-driven \texttt{for-in} loop consistent
with the \texttt{Sequence}-driven case, since the
compiler computes the type of ``one iteration's worth'' of 
the expansion and then checks that the loop pattern can
bind such a value. The resulting 
\texttt{PackIterationInfo} records the pattern type so
that it can be later rewritten and simplified after solving.

The constraint application (\texttt{CSApply})
mirrors this approach, with a special code path
introduced for the value pack-driven \texttt{for-in} loop.
The existing \texttt{applySolutionToForEachStmt}
logic was similarly split into sequence and
pack overloads, with a dispatcher
selecting the appropriate path based on whether
the stored loop information is \texttt{SequenceIterationInfo} or \texttt{PackIterationInfo}. For the new pack
case, the solution application performs
three conceptually distinct tasks:

\begin{enumerate}
    \item It rewrites the pack expansion target
using the solved constraints.
    \item It simplifies the stored pattern type,
removing any remaining type variables introduced during
constraint generation.
    \item It propagates the opened element environment
    into the declarations
    that appear in the loop pattern and
    body. \label{itm:third}
\end{enumerate} 

We implemented Step 3 using a dedicated AST walker over the 
loop pattern and body. Whenever the
walker encounters a \texttt{VarDecl}, it records the
generic environment obtained from the pack expansion by
calling \texttt{setOpenedElementEnvironment}.

Function and nominal declarations are not recursively
rewritten in the same way, because the goal is to annotate the declarations in the scope of the current pack
iteration, not to rebind nested declaration contexts.

The effect is that local bindings inside the loop
are type-checked as if they had been introduced in a
scalar generic context corresponding to one opened pack
element. Figure~\ref{fig:pack-element-mapping} illustrates this
typing model for a call whose argument pack has element types
\texttt{\{Int, String, Bool\}}. Although the loop iterates over
\texttt{repeat each ts}, the body does not type-check \texttt{t}
as a pack. Instead, the opened element environment gives
\texttt{t} a scalar element type, written in the figure as
\texttt{T'}, which is bound to \texttt{Int}, \texttt{String},
and \texttt{Bool} at the corresponding conceptual pack iterations.

The opened element environment also explains why
pack iteration is more subtle than just allowing a new
expression in loop position. In ordinary inline pack
expansion expression, the pattern is checked in a
context where there is no local variable,
and the expression is applied to the entire pack.
In pack iteration, however, the body must be checked in
a context where those same pack references have been
opened into scalar element types. This
distinction becomes visible in later extensions as well:
for example, support for a \texttt{where} clause
is not trivial precisely because the
\texttt{where} condition is currently type-checked
together with the loop pattern, while the
opened element environment needed for
scalar type-checking is only available 
after the pattern has been processed.

\begin{figure}[t]
  \centering
  \begin{tabular}{@{}c@{\hspace{1.0em}}c@{}}
    \begin{minipage}[t]{0.55\columnwidth}
\begin{lstlisting}
func iterate<each T>(
  _ ts: repeat each T
) {
  for t in repeat each ts {
    // t has scalar type T'
    use(t)
  }
}

iterate(1, "hello", false)
\end{lstlisting}
    \end{minipage}
    &
    \begin{minipage}[t]{0.38\columnwidth}
      \centering
      \scriptsize
      \[
      \begin{array}{c|c|c}
        i & \texttt{T'} & \texttt{t} \\
        \hline
        0 & \texttt{Int}    & \texttt{1} \\
        1 & \texttt{String} & \texttt{"hello"} \\
        2 & \texttt{Bool}   & \texttt{false}
      \end{array}
      \]
    \end{minipage}
  \end{tabular}

  \caption{Opening a type pack into scalar element contexts during
  pack iteration. Given a value pack \texttt{ts} with heterogeneous
  element types, each loop iteration binds \texttt{t} as a local
  value of a scalar generic type \texttt{T'}.}
  \label{fig:pack-element-mapping}
\end{figure}

\subsubsection{Nested Iteration}

One further consequence of this design is that
nested pack iteration requires the
compiler to retain outer opened element
environments while type-checking inner
loops. In the implementation, this is handled
by caching a stack of pack-element
generic environments in the constraint system and
threading the current environment through nested
\texttt{for-in repeat} targets. This allows inner loop
bodies to refer both to their own current
element and to values bound by
enclosing pack iterations without losing
the scalar interpretation of either context.

Consider the following function, where the inner pack
loop refers to the element bound by the outer loop:

\begin{lstlisting}
func nested<each T, each U>(
  value: repeat each T,
  value1: repeat each U
) {
  for e1 in repeat each value {
    for _ in [] {}
    
    for e2 in repeat each value1 {
      let y = e1
      _ = e2
      _ = y
    }

    let x = e1
    _ = x
  }
}
\end{lstlisting}

The empty \texttt{Sequence} loop on line 6 is included only to show
that ordinary loops do not affect the environment stack described below.
Note that in the above code, \texttt{nested} can be called
with value packs of completely different shapes and sizes, and
the compiler needs to be able to infer the type of the local variables
\texttt{e1} and \texttt{e2} correctly.

Let $E_T$ denote the opened element environment for the outer loop over
\texttt{value}, and let $E_U$ denote the opened element environment
for the inner loop over \texttt{value1}. Under our implementation,
nested pack iteration is type-checked by maintaining a stack of
pack-element environments, as traced in Table~\ref{tab:nested-env}. Entering a pack loop pushes its opened element environment, and escaping the scope of that loop pops it. Ordinary 
\texttt{Sequence}-based loops do not affect the stack.

\begin{table}[t]
\centering
\caption{Opened element environment stack while type-checking
\texttt{nested}. Line numbers refer to the \texttt{nested} listing.}
\label{tab:nested-env}
\begin{tabular}{p{0.12\linewidth}p{0.50\linewidth}p{0.24\linewidth}}

\toprule

Lines & Event & Environment \\

\midrule

1--4   & Enter function body                                      & $[\ ]$ \\

5      & Enter outer pack loop, bind \texttt{e1}                  & $[E_T]$ \\

6      & Enter \texttt{Sequence} loop                    & $[E_T]$ \\

6      & Exit \texttt{Sequence} loop                     & $[E_T]$ \\

8      & Enter inner pack loop, bind \texttt{e2}                  & $[E_T, E_U]$ \\

9--11  & Type-check inner loop body (\texttt{y}, \texttt{e2})     & $[E_T, E_U]$ \\

12     & Exit inner pack loop                                     & $[E_T]$ \\

14--15 & Type-check outer loop body after inner loop (\texttt{x}) & $[E_T]$ \\

16     & Exit outer pack loop                                     & $[\ ]$ \\

\bottomrule

\end{tabular}
\end{table}

The stack ensures that the innermost pack loop is interpreted in its
own scalar element context, while still preserving access to enclosing pack
contexts. This is why \texttt{e1} remains usable both inside
the inner pack loop and after that loop's scope has ended.

\subsubsection{SIL Generation}

The SILGen pass of the Swift compiler
lowers the type-checked abstract syntax tree
obtained in the Semantic Analysis step down to raw,
unoptimized Swift Intermediate Language (SIL) \cite{swift-sil-doc}. 

After the raw SIL is obtained, the compiler then
optimizes it and produces LLVM IR \cite{Lattner2004LLVM}, which then generates
machine code for various processor architectures.

Before \emph{Pack Iteration}, given a pack expansion 
expression, it was possible to emit
a dynamic pack loop using the existing\linebreak
\texttt{emitDynamicPackLoop} function.
This function was used by SILGen to generate SIL for an inline
pack expansion expression.

We modified the existing lowering of the \texttt{for-in} loop in SILGen to call this function when the source of the loop
is a pack expansion expression. More concretely, the lowered SIL maintains
an iteration index, checks it against the pack length, opens the current pack
element, binds it to the loop pattern, executes the body, and only then advances to the next element.

This modification enabled the compiler to emit valid SIL for a trivial pack iteration case such as:

\begin{lstlisting}
func iterate<each Element>(
  over element: repeat each Element
) {
  for el in repeat each element {}
}
\end{lstlisting}

For the above code, the emitted SIL has the following simplified
shape:

\begin{lstlisting}[language=]
bb0:
  %start = 0
  %step = 1
  %length = pack_length ...
  br bb1(%start)

bb1(%i):
  %done = cmp_eq_Word(%i, %length)
  cond_br %done, bb2, bb3

bb2:
  return

bb3:
  %packIndex = dynamic_pack_index %i ...
  %elt = open_pack_element %packIndex ...
  bind loop pattern
  execute loop body
  %next = add_Word(%i, %step)
  br bb1(%next)
\end{lstlisting}

We have added the support for
\texttt{break} and \texttt{continue} statements by
recording their destinations in the same data structure
used by the existing \texttt{for-in} loop.
Consequently, neither construct required a
pack-specific lowering rule. A \texttt{continue}
transfers control to the loop's continuation destination,
which rejoins the back-edge of the dynamic pack loop,
whereas a \texttt{break} transfers control to the exit destination
emitted after the loop.

However, using the existing lowering primitive has proven 
insufficient for supporting
more advanced usage of \texttt{for-in} loop in Swift, such
as pattern-matching with \texttt{for case} statement:

\begin{lstlisting}
enum E<T> {
  case one(T)
  case two
}

func iteratePatternMatch<each Element>(
  over element: repeat E<each Element>
) {
  for case .one(let value) in repeat each element {
    print(value)
  }
}

iteratePatternMatch(over: E<Int>.one(5), E<Int>.two)
// Prints '5'
\end{lstlisting}

In the program above, the pattern-matching with
\texttt{for case} in the loop statement ensures that
the body of the loop is only evaluated for the first
element of the value pack passed as argument: \texttt{E<Int>.one(5)}.

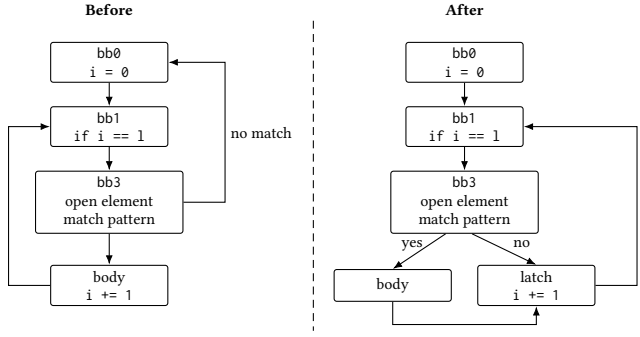
\begin{figure}[t]
\centering
\begin{tikzpicture}[
  font=\scriptsize,
  block/.style={
    draw,
    rounded corners=1pt,
    minimum width=1.55cm,
    minimum height=0.42cm,
    align=center
  },
  wideblock/.style={
    draw,
    rounded corners=1pt,
    minimum width=1.95cm,
    minimum height=0.48cm,
    align=center
  },
  arrow/.style={-{Latex[length=1.25mm]}, thin},
  every node/.style={inner sep=2pt}
]

\node[font=\bfseries\scriptsize] at (0,0) {Before};
\node[font=\bfseries\scriptsize] at (4.7,0) {After};

\node[block]     (b0)    at (0,-0.70) {\texttt{bb0}\\\texttt{i = 0}};
\node[block]     (b1)    at (0,-1.55) {\texttt{bb1}\\\texttt{if i == l}};
\node[wideblock] (b3)    at (0,-2.55) {\texttt{bb3}\\open element\\match pattern};
\node[block]     (bbody) at (0,-3.65) {body\\\texttt{i += 1}};

\draw[arrow] (b0) -- (b1);
\draw[arrow] (b1) -- (b3);
\draw[arrow] (b3) -- (bbody);

\draw[arrow]
  (bbody.west) -- ++(-0.55,0)
  |- (b1.west);

\draw[arrow]
  (b3.east) -- ++(0.55,0)
  |- node[pos=0.25, right] {no match} (b0.east);

\node[align=center, text width=2.30cm] at (0,-4.45) {};

\node[block]     (a0)     at (4.7,-0.70) {\texttt{bb0}\\\texttt{i = 0}};
\node[block]     (a1)     at (4.7,-1.55) {\texttt{bb1}\\\texttt{if i == l}};
\node[wideblock] (a3)     at (4.7,-2.55) {\texttt{bb3}\\open element\\match pattern};
\node[block]     (abody)  at (3.75,-3.65) {body};
\node[block]     (alatch) at (5.65,-3.65) {latch\\\texttt{i += 1}};

\draw[arrow] (a0) -- (a1);
\draw[arrow] (a1) -- (a3);

\draw[arrow]
  ([xshift=-0.25cm]a3.south) --
  node[pos=0.32, left] {yes}
  (abody.north);

\draw[arrow]
  ([xshift=0.10cm]a3.south) --
  node[pos=0.32, right, xshift=0.2cm] {no}
  (alatch.north);

\draw[arrow]
  (abody.south) -- ++(0,-0.30)
  -| (alatch.south);

\draw[arrow]
  (alatch.east) -- ++(0.55,0)
  |- (a1.east);

\node[align=center, text width=2.30cm] at (5.65,-4.45) {};

\draw[densely dashed] (2.7,-0.15) -- (2.7,-4.25);

\end{tikzpicture}

\caption{Control-flow shape for pack iteration before and after
introducing a loop latch. The latch factors index advancement into a
shared continuation reached by both successful body execution and
pattern-match failure.}
\label{fig:loop-latch}
\end{figure}

For this program, the emitted SIL would match the first element,
but the second one would be skipped, branching
back to the loop setup block rather than to the loop continuation.
Because the setup block initializes the dynamic pack index to zero,
it would restart the iteration from the beginning of the pack instead of
advancing to the next element, resulting in an infinite loop.
Figure~\ref{fig:loop-latch} illustrates this control-flow problem.

To solve this problem, we added a ``loop latch,'' a separate block
responsible for advancing the dynamic pack index:

\begin{lstlisting}[language=]
bb6:
  %next = add_Word(%i, %step)
  br bb1(%next)
\end{lstlisting}

Both normal body
evaluation and pattern-match failure branch to the latch before
returning to the loop header. As such, the loop body evaluation
block \texttt{bb3} was no longer responsible for advancing the index.

This completes the lowering of parameter-pack iteration, 
allowing the Swift compiler to perform the necessary 
optimizations and translate the optimized SIL into LLVM IR.

The Swift compiler changes in semantic analysis and SIL
lowering made value parameter packs in Swift a legitimate
source for a \texttt{for-in} loop and, in effect, allow the
programmer to reason about an advanced feature of
variadic generics by using them in a familiar
construct. This makes parameter packs in Swift approachable
and familiar, enabling programmers to introduce them
to solve problems that were previously impossible with scalar
generics in their code.

\section{Evaluation}

In this section, we detail the practical application of
pack iteration and the patterns Pack Iteration enables for Swift programmers,
performance observations, limitations, and comparison with emerging
C++26 features.

\subsection{New Functionality}
\label{sec:newfunc}

Pack iteration enables programmers to effortlessly
introduce variadic generics into their code while making
it possible to express new types of operations in a performant 
and readable way.

Let us consider the following examples that demonstrate this claim.

\begin{lstlisting}
func allEmpty<each T>(
    _ array: repeat [each T]
) -> Bool {
  for a in repeat each array {
    guard a.isEmpty else { return false }
  }

  return true
}

print(allEmpty(["One", "Two"], [1], [true, false], []))
// false
\end{lstlisting}

The above construction allows the Swift programmer to operate
on a pack of heterogeneously-typed arrays, and in a succinct
way check that all of them are empty.

By using the on-demand evaluation property of pack iteration, the
programmer is able to short-circuit and exit the function early
when the first non-empty array is encountered.

The \texttt{ValueProducer}/\allowbreak\texttt{evaluateAll} example from Section \ref{subsec:design} further demonstrates this new functionality.

\subsection{Performance}

To analyze the performance of \emph{Pack Iteration}, we designed four
benchmarks targeting common patterns programmers would employ when
using this feature.

Each benchmark has two versions. The first, referred to as ``Before'', uses pack expansion expressions together with a local
helper function. The second, referred to as ``After'', uses pack
iteration. The two versions are logically equivalent.

The benchmarks are as follows:
\begin{enumerate}
  \item \emph{Tuple} measures tuple equality over two heterogeneous
  value packs. The ``Before'' version simulates early exit using a
  local throwing helper function, while the ``After'' version uses a
  \texttt{for-in repeat} loop with an ordinary \texttt{guard} statement.

  \item \emph{Lookup} measures a simple generic-requirement lookup
  pattern. Each pack element conforms to a protocol, and the benchmark
  calls a protocol requirement on every element. This tests whether pack
  iteration introduces overhead when the loop body performs ordinary
  generic member lookup.

  \item \emph{Result} measures
  pattern-based processing over a pack of enum values. Similarly
  to the pattern described in Section~\ref{sec:newfunc}, this benchmark
  processes only the \texttt{.success} cases of a pack of
  \texttt{Result<Success, Failure>} values, comparing a helper-function-and-\texttt{switch}
  implementation with a direct \texttt{for case} pack iteration loop.

  \item \emph{Closure} measures pack processing when each element is
  handled through closure-based control flow. This benchmark represents
  code where the cost of closure capture and indirect invocation may
  dominate the cost of the pack-processing mechanism itself.
\end{enumerate}

For each benchmark, we measured compile time, run time, and binary size
as the number of pack elements increased from 100 to 1000. Each
measurement was averaged over 10 runs.
We emphasize that scaling pack sizes from 100 to 1,000 elements is not intended to model typical application workloads, where tuples and packs rarely exceed a few elements. Rather, this range serves as an asymptotic stress test to evaluate the scalability, solver memory overhead, and linear bounds of the Constraint System (CSGen/CSApply) and SIL lowering under extreme compilation loads.

\begin{figure}[t]
  \centering
  \includegraphics[width=\columnwidth]{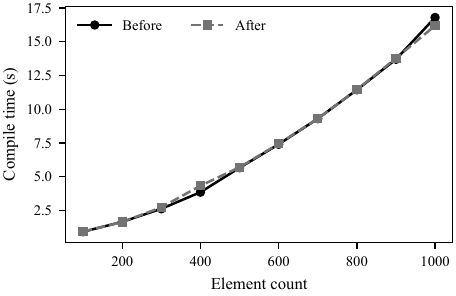}
  \caption{Compile time for the \emph{Result} benchmark before
  and after pack iteration.}
  \label{fig:result-compile-time}
\end{figure}

\begin{figure}[t]
  \centering
  \includegraphics[width=\columnwidth]{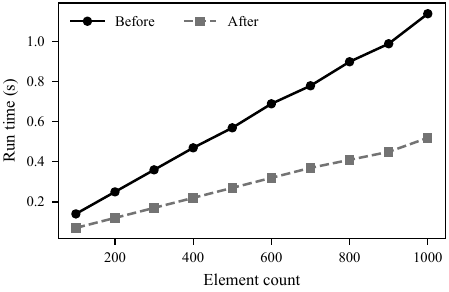}
  \caption{Run time for the \emph{Result} benchmark before
  and after pack iteration.}
  \label{fig:result-runtime}
\end{figure}

\begin{figure}[t]
  \centering
  \includegraphics[width=\columnwidth]{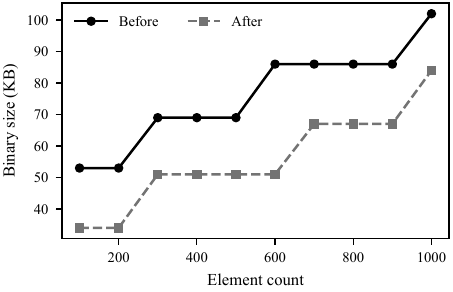}
  \caption{Binary size for the \emph{Tuple} benchmark before and after pack
  iteration.}
  \label{fig:tuple-binary-size}
\end{figure}

Across the benchmarks, pack iteration does not introduce a
performance penalty. As summarized in
Table~\ref{tab:benchmark-summary}, compile times remain
comparable between the ``Before'' and ``After''
versions at 1000 pack elements. The largest compile-time
differences are small: the \emph{Tuple} and \emph{Closure} benchmarks
slow down by 1.8\%, while \emph{Lookup} and
\emph{Result} speed up by 2.2\% and 3.6\%, respectively. Figure~\ref{fig:result-compile-time}
shows this trend for the \emph{Result} benchmark
across the full range of measured pack sizes.

The more interesting effects appear at run time and
in code size, particularly in benchmarks where the
``Before'' version relied on additional control-flow
workarounds such as throwing errors or helper
functions.

Figure~\ref{fig:result-runtime} shows the clearest
run time improvement, for the \emph{Result} 
benchmark. At 1000 
elements, run time decreases from 1.14\,s to 0.52\,s, a
reduction of 54.4\%, as shown in Table~\ref{tab:benchmark-summary}.
This improvement is consistent with the structure of
the two programs. Pack iteration removes an extra layer
of per-element dispatch introduced by the local generic
helper function and replaces it with a loop whose control
flow is explicit in the source code and in the generated
SIL. Therefore, the resulting code reduces the per-element run-time overhead introduced by existentials in Swift~\cite{Barik2019SwiftProtocols, Racordon2025Existentialize}.

The \emph{Closure} benchmark results show a marginal improvement of only $0.2\%$. This suggests that the substantial runtime overhead inherent to Swift's closure capture and indirect dispatch masks the efficiencies gained through loop lowering. While pack iteration improves readability in these contexts, the performance benefits are bottlenecked by the underlying closure implementation rather than the iteration mechanism itself.

Figure~\ref{fig:tuple-binary-size} shows the most
pronounced code-size improvement, for tuple equality.
Here, the ``Before'' version must simulate short-circuiting
by introducing an auxiliary \texttt{Error} type, a local
throwing helper, and a surrounding \texttt{do}/\texttt{catch}
block. The ``After'' version expresses the same logic
directly with \texttt{for (left, right) in repeat} loop
and an ordinary \texttt{guard} in the loop's body. At 1000 
elements, binary size drops from 102\,KB to 84\,KB, a
reduction of 17.6\%, as summarized in 
Table~\ref{tab:benchmark-summary}.

This encouraging result stems from all the extra code required before
pack iteration to simulate the short-circuiting
behavior, increasing the total code size of the program.

Taken together, these benchmarks show that pack iteration
not only makes pack-processing code more direct and
idiomatic, but does so without imposing a performance
penalty. In cases where the older workaround-based
style required helper functions, throwing control flow, or
other auxiliary machinery, the direct iteration form is not
merely easier to read; it also produces code that is
smaller and more performant.

\begin{table}[t]
\centering
\scriptsize
\caption{Summary of benchmark results at 1000 pack elements. Compile time and run time are reported in seconds. Binary size is reported in KB.}
\label{tab:benchmark-summary}
\begin{tabular}{llrrr}
\toprule
Benchmark & Metric & Before & After & Change \\
\midrule
Tuple
  & Compile time & 264.88 & 269.69 & +1.8\% \\
  & Run time     & 0.22   & 0.21   & -4.5\% \\
  & Binary size  & 102    & 84     & -17.6\% \\
\midrule
Lookup
  & Compile time & 13.48 & 13.19 & -2.2\% \\
  & Run time     & 0.09  & 0.06  & -33.3\% \\
  & Binary size  & 51    & 51    & 0.0\% \\
\midrule
Result
  & Compile time & 16.80 & 16.20 & -3.6\% \\
  & Run time     & 1.14  & 0.52  & -54.4\% \\
  & Binary size  & 137   & 136   & -0.7\% \\
\midrule
Closure
  & Compile time & 172.70 & 175.88 & +1.8\% \\
  & Run time     & 221.00 & 220.61 & -0.2\% \\
  & Binary size  & 192    & 187    & -2.6\% \\
\bottomrule
\end{tabular}
\end{table}

\subsection{Comparison to Emerging C++ and Rust Features}

C++ is also evolving toward more direct mechanisms for
operating on parameter packs. Two emerging C++26 features are
especially relevant: pack indexing and expansion
statements~\cite{Jabot2023PackIndexing, ExpansionStmts}.
These features address the same broad usability problem that
motivates Swift pack iteration: programmers
should not need to encode simple pack-processing operations through 
recursive templates, helper metafunctions, or lambda-based library workarounds.

Pack indexing provides direct access to an individual element of a
parameter pack by index. This improves element selection, since
the programmer can now refer to a specific pack element without
recursively decomposing the pack.
However, to process every element with pack indexing, the programmer
must still combine indexing with some other mechanism for
generating the indices.

Expansion statements are more directly
comparable to pack iteration in Swift. The proposal introduces
\texttt{template for}, a statement form that enables compile-time
repetition of a statement over expression lists, tuple-like objects,
ranges with compile-time size, and types via reflection
\cite{ExpansionStmts}. For example, a C++ expansion statement over a pack expression list has the following form:

\begin{lstlisting}[language=C++]
template <typename... Ts>
void print_all(Ts... elems) {
  template for (auto elem : {elems...}) {
    std::println("{}", elem);
  }
}
\end{lstlisting}

The proposal specifies this construct by determining an expansion size
and expanding the body into one statement block per element. In the
example above, the body is expanded at compile-time into separate blocks binding
\texttt{elem} to \texttt{elems...[0]}, then \texttt{elems...[1]}, and so
on.
The transition from recursive template decomposition to an imperative loop structure represents more than a syntactic convenience; it is a move toward making the ``machinery'' of the language a first-class, accessible abstraction. This follows the historical precedent of creating more ``pliable'' meta-programming interfaces that reduce the cognitive load on the developer~\cite{Kiczalesetal1991}.

Nevertheless, this model is distinct from the approach presented in this paper,
since rather than expanding the loop body into a fixed sequence of statement
blocks, Swift lowers a pack expansion used as a \texttt{for-in} source
into a dynamic pack loop. The compiler manipulates the iteration index,
keeps track of the current pack element environment, executes
the loop body, and then advances to the next element. Thus, where C++
expansion statements make pack processing look like a loop while
remaining fundamentally a compile-time expansion mechanism, pack iteration in
Swift makes a  value pack the first-class loop source.

This distinction matters for both execution and implementation. In
Swift, pack iteration has an on-demand execution model: later pack
elements need not be opened or evaluated if control flow exits the loop
early. Ordinary control flow constructs such as \texttt{break},
\texttt{continue}, and early \texttt{return} are represented directly in
the loop-shaped SIL control flow. C++ expansion statements also define
the behavior of \texttt{break} and \texttt{continue}, but they do so over
a compile-time expanded sequence of statement blocks. Swift's
implementation instead reuses the compiler's ordinary loop-control
machinery by lowering the feature as a dynamic loop over pack elements.

This distinction is rooted in the differences of the generic models
in the two languages, as discussed in Section \ref{sec:variadicC++}.
C++ variadic templates are instantiated at compile-time,
so pack manipulation naturally appears as part of compile-time metaprogramming. 
Swift generics, by contrast, support
separately type-checked generic declarations and unspecialized generic
code using runtime type metadata. This design lets Swift express
pack iteration with an ordinary \texttt{for-in} loop, where the
local variable is treated as a regular value of a scalar generic
type.

Rust language-design discussions point in a related direction. Rust does not
currently support variadic generics as a language feature, but design notes 
and pre-RFC discussions have repeatedly explored tuple-based models of variadics 
and explicit mechanisms for expanding or iterating over tuple 
elements~\cite{RustVariadicGenericsDesignNotes, RustVariadicGenericsSketch}. Some of these discussions propose constructs
such as \texttt{static for}, which would make operations over variadic structure 
look more like ordinary iteration while preserving Rust's compile-time type reasoning.

The emergence of similar features in C++26 and Rust suggests a broader trend in programming languages, where previously advanced features such as parameter packs are being made more approachable and composable with constructs that programmers already understand.
However, while C++26 expansion statements provide a syntactic loop, they remain fundamentally a compile-time unrolling mechanism. Rust's proposed designs similarly emphasize compile-time reasoning over tuple-like variadic structure.
Swift instead lowers the pack to a sequential loop, so programmers can reason about processing a pack the same way they reason about iterating over an array.
As a result, Pack Iteration turns what was an expert-only feature into one that fits naturally into ordinary Swift code.

\subsection{Community Review and Language Evolution}

Swift language changes are developed through the Swift Evolution process, a public proposal and review process centered around the Swift
Forums.\footnote{\url{https://forums.swift.org/}} The forums
serve as the main venue where new language features are
pitched, discussed, and formally reviewed before acceptance.
Notably, Swift's review process does not evaluate proposals 
only as isolated solutions to technical problems.
It asks whether a given proposal fits the ``feel and direction'' of Swift, composing well with the existing features
and upholding the ``progressive disclosure'' paradigm, where
advanced capabilities are exposed through constructs that remain approachable at first use and reveal their
full power as programmers encounter more demanding use cases~\cite{SwiftEvolutionProcess}.

Community feedback on the Swift Forums was overwhelmingly
positive and consistently framed \emph{Pack Iteration}
as the missing usability layer for variadic generics.
Swift language community members have repeatedly
described the feature as ``intuitive,'' ``natural,''
and necessary for making packs practical in day-to-day
Swift code.\footnote{\url{https://forums.swift.org/t/review-se-0408-pack-iteration/67152}}

This process also provides an instructive contrast with C++. 
Related C++26 features such as pack indexing and expansion 
statements were developed through WG21, the ISO C++
standards committee, using numbered papers, working-group
review, and consensus among accredited national-body
experts~\cite{WG21Committee,WG21Practices}. This process
must account for a multi-decade language, multiple independent implementations, and the constraints of
international standardization.

Both communities are responding to the same pressure to make variadic generics more usable, but they channel that
pressure differently: Swift Evolution emphasizes public
community review, implementation experience, and language
feel, whereas WG21 emphasizes standardizable specification,
multi-implementation viability, and committee consensus.

\subsection{Limitations and Future Work}

Today, \emph{Pack Iteration} as presented in this paper
still lacks two main features that are natural in a context
of regular \texttt{for-in} loops in Swift over \texttt{Sequence} types, but
are nontrivial to implement when a value pack is
the loop source and are considered a future direction.
Let us enumerate these features next.

\subsubsection{Support of \texttt{where} Clause}

One of the features supported by an ordinary
\texttt{for-in} loop over a \texttt{Sequence}
type is the \texttt{where} clause. The \texttt{where}
clause in Swift lets the programmer evaluate
only those elements that satisfy a given condition. For example:

\begin{lstlisting}
func iterateOverNonempty<each T: Collection>(
  ts: repeat each T
) {
  for t in repeat each ts where !t.isEmpty {
    // code
  }
}
\end{lstlisting}

In the above code, only the non-empty collections
will be evaluated.

To support this feature, simply going through the codepath
that the \texttt{Sequence}-based \texttt{for-in} loop
takes alone would not be sufficient. The limitation of this
approach is that in the current design of the \texttt{for-in}
loop, the \texttt{where} clause is type-checked
together with its pattern.

In the case of pack iteration, this poses a problem because
the opened element environment necessary for type-checking
the statement following the \texttt{where} clause can only
be obtained after the pattern is type-checked. So, this
approach will introduce an ordering dependency, where the
opened element environment may or may not be ready
when the \texttt{where} clause is type-checked.

A future direction, therefore, would be to
reimplement the existing logic to defer type-checking of the \texttt{where} clause until the pattern environment is fully established.

\subsubsection{Single-lane Assignment}

Given the changes presented in this paper, it is not
possible to assign into a single
lane of a pack inside of a \texttt{for-in-repeat}
loop. For example:

\begin{lstlisting}
func wrap<each T>(
  _ values: repeat each T
) -> (repeat Optional<each T>) {
  var result: (repeat Optional<each T>)
  for value in repeat each values {
    // initialize the corresponding lane of `result`
    // result[current lane] = .some(value)
  }
  return result
}
\end{lstlisting}

Supporting initialization of a single element of a
tuple such as \texttt{result} would unlock transformations like the one above, where a new tuple of the same shape is built, element by element, from the
corresponding elements of a value pack within
the loop body. Since the compiler statically knows
the shape of the iteration, this restriction may be lifted
by a future extension to definite initialization.

\section{Conclusion}

Variadic generics are powerful, but often unintuitive to use. In many
languages, operations over parameter packs require programmers to reach for
nontrivial techniques, such as recursive template decomposition,
helper functions, or expression-level expansion patterns that do not compose
naturally with ordinary control flow.

This paper presented \emph{Pack Iteration}, a language feature that makes
parameter packs usable through a familiar \texttt{for-in} loop in Swift.
Under these changes, a pack expansion expression can now serve as the
source of iteration, and each iteration opens one element of the pack into
a scalar generic context. This lets programmers bind pack elements to local
variables, use pattern matching, and write familiar control-flow constructs
such as \texttt{break}, \texttt{continue}, and early \texttt{return}.

The implementation required changes across the Swift compiler. In semantic
analysis, the compiler distinguishes sequence iteration from pack iteration,
derives the scalar element type from the expansion pattern, and records the
opened element environment needed to type-check declarations inside the loop.
In SILGen, the compiler lowers pack iteration to a dynamic pack loop that
opens and evaluates elements on demand, rather than eagerly expanding the
entire pack. This design preserves Swift's model of separately type-checked
generic code while giving parameter packs loop-like behavior.

Our evaluation shows that pack iteration improves the expressiveness and
readability of variadic-generic code without imposing a performance penalty.
In several benchmarks, the direct iteration form also improves run time
(up to $54.4\%$ reduction) or binary size (up to $17.6\%$ reduction) by
eliminating workaround patterns such as local helper functions and
throwing control flow.

This work also reflects a broader direction in language design. Swift is
not alone in recognizing that variadic generics should compose with
ordinary programming constructs rather than remain confined to
special-purpose metaprogramming idioms. C++26 expansion statements, for
example, introduce a \texttt{template for} 
statement that brings loop-like syntax to
compile-time expansion. Rust language design discussions have similarly 
considered tuple-centered variadic 
generics, including proposals with 
explicit iteration or expansion constructs 
such as \texttt{static for}. These efforts differ in
their generic models and implementation strategies, but they point to
the same underlying pressure: programmers need direct, readable ways to
operate over heterogeneous sequences of types and values.

More generally, pack iteration shows that variadic generics need not
remain an expert-only feature. Swift's contribution is one point in a
larger design space: it integrates parameter packs with a familiar
\texttt{for-in} loop while preserving static type information,
on-demand execution, and compatibility with Swift's separately
type-checked generic implementation model. As other languages explore
related mechanisms, pack iteration provides evidence that advanced
generic programming can be made more approachable without giving up type
safety or efficient execution.

\newpage{}

\begin{acks}
We would like to thank the members of the Swift Core Team,
in particular, Holly Borla, Anthony Latsis, and Michael Gottesman,
for providing invaluable feedback and guidance for the design and implementation of
the parameter pack feature.
\end{acks}

\bibliographystyle{ACM-Reference-Format}
\bibliography{packs}

\end{document}